\documentclass[english,aps,manuscript,reprint, twocolumn,superscriptaddressm,showkeys]{revtex4-1}
\usepackage{eurosym}
\usepackage{graphicx}
\usepackage{dcolumn}
\usepackage{bm}
\usepackage{color}
\usepackage{latexsym}
\usepackage{amssymb}
\usepackage{amsmath}

\begin{document}
	
	\title{Optimal leap angle of legged and legless insects in a landscape of uniformly-distributed random obstacles}
	\author{Fabio Giavazzi}
	\affiliation{Dipartimento di Biotecnologie Mediche e Medicina Traslazionale, Universit\`a degli Studi di Milano, I-20133 Milano, Italy} 

	\author{Samuele Spini}
	\affiliation{Dipartimento di Fisica A. Pontremoli, Universit\`a degli Studi di Milano, I-20133 Milano, Italy}
	
	\author{Marina Carpineti}
	\affiliation{Dipartimento di Fisica A. Pontremoli, Universit\`a degli Studi di Milano, I-20133 Milano, Italy}
	
	\author{Alberto Vailati}
	\email{corresponding author: alberto.vailati@unimi.it}
	\affiliation{Dipartimento di Fisica A. Pontremoli, Universit\`a degli Studi di Milano, I-20133 Milano, Italy}

	\begin{abstract}
		\textbf{We investigate theoretically the ballistic motion of small legged insects and legless larvae after a jump. Notwithstanding their completely different morphologies and jumping strategies, these legged and legless animals have convergently evolved to jump with a take-off angle of 60$^\circ$, which differs significantly from the leap angle of 45$^\circ$ that allows reaching maximum range. We show that in the presence of uniformly-distributed random obstacles the probability of a successful jump is directly proportional to the area under the trajectory. In the presence of negligible air drag, the probability is maximized by a take-off angle of 60$^\circ$. 
			The numerical calculation of the trajectories shows that they are significantly affected by air drag, but the maximum probability of a successful jump still occurs for a take-off angle of 59-60$^\circ$ in a wide range of the dimensionless Reynolds and Froude numbers that control the process.  We discuss the implications of our results for the exploration of unknown environments such as planets and disaster scenarios by using jumping robots.}
	\end{abstract}

	\maketitle
	\section{Introduction\label{sec:Intro}}
	Some animal species have evolved the ability of jumping, both as a fast locomotion method and as an escape maneuver \cite{biewener2018}.  Compared to deambulation or locomotion by crawling, jumping allows achieving a fast-displacement by simultaneously overcoming natural obstacles distributed across the landscape \cite{irschick2016}. The evolutive process leading to the optimization of the jumping performances is strongly influenced by the features of the habitat, which contribute to determining the selection of a particular take-off angle in an animal species \cite{Toro2004}. 
	
	Jumping performances become remarkable in insects that use a catapult mechanism to amplify their muscular power and achieve long-range jumps by storing elastic energy inside their exoskeleton \cite{Patek2011}. When the elastic energy is suddenly released by unlocking a latch, the body is projected into the air at high speed at distances significantly larger than the size of their body. Insects have devised drastically different methods to jump by using a catapult mechanism. Froghoppers (\textit{Philaenus spumarius}) take advantage of elastic deformation of their chitinous exoskeleton to achieve a rapid extension of their hind legs \cite{Burrows2008}. 
	As a result, the body undergoes an acceleration as large as 400 m/s$^2$  and enters a ballistic phase with top speeds up to 4 m/s and a typical take-off angle of 58$^\circ \pm 2.6^\circ$ \cite{Burrows2003} (Table \ref{tab:jump}).

	Jumping driven by a catapult mechanism can be achieved also in the absence of highly specialized limbs like those present in froghoppers. A remarkable example is represented by the legless jumping of the larvae of the Mediterranean fruit fly (\textit{Ceratitis capitata}),  which take advantage of a peculiar hydrostatic catapult mechanism that allows them to jump to distances of the order of 12 cm, namely more than ten times their body length, in a fraction of a second \cite{Maitland1992}. 
	In this case, a hydrostatic skeleton made by flexible outer muscular bands is anchored to a flexible skin layer, which confines an inner fluid region \cite{Kier2012}. To jump, the larva first contracts its longitudinal ventral muscles to form a loop and anchors its head to the tail by using a pair of mouth hooks. The inner pressure of the body is then increased by contraction of helical muscular bands, and the sudden release of the mouth hooks gives rise to an abrupt straightening of the body, leading to a rapid acceleration phase ending with take-off at a velocity of about 1.2 m$/$s and a take-off angle of 60$^\circ$ \cite{Maitland1992} (Table \ref{tab:jump}).
	A similar hydrostatic catapult mechanism is adopted by the larvae of gall midge (\textit{Asphondylia sp.}) \cite{Farley2019}. In this case, latching is achieved by micrometer-scale finger-like microstructures distributed across the width of body segments. The sudden release of this latch leads to a take-off velocity of 0.85 m/s at a take-off-angle of $63^\circ \pm 3^\circ$ (Table \ref{tab:jump}).

	\begin{table*}[t!]
		\begin{center}
			\begin{tabular}{c|c|c|c|c|c|c|c} 
				& \textbf{mass}& \textbf{body} & \textbf{effective}& \textbf{take-off}&\textbf{take-off} & \textbf{substrate} & \textbf{Ref.}\\
				
				\textbf{Species} & ($\times 10^{-6}$kg)&\textbf{length}&\textbf{diameter}&\textbf{speed}&\textbf{angle} & \textbf{} & \textbf{}\\
				
				& & ($\times 10^{-3}$m) & ($\times 10^{-3}$m)& (m/s)& &  & \\
				
				\hline 
				& &   & &&&&\\
				Gall midge larva&1.27&3.28&1.2&0.85&$63.3 \pm3^\circ$ &plastic & \cite{Farley2019}\\
				\textit{(Asphondylia sp.)}&&&&&  &   & \\
				& &  & &&&&\\
				
				Fruit fly larva	&17&8.5&2.8&1.17&$60^\circ$ &- & \cite{Maitland1992}\\
				\textit{(Ceratitis capitata)}&&&&	&  &  & \\
				& &   & &&&&\\
				Froghopper & 12.3&6&4&4& $58\pm2.6^\circ$ & - & \cite{Burrows2003}\\
				(\textit{Philaenus spumarius})&&&&&$53.2\pm13^\circ$ &epoxy  & \cite{Goetzke2019}\\
				&&&&&$53.6\pm14^\circ$ &ivy leave  & \cite{Goetzke2019}\\
			\end{tabular}
		\end{center}
		\caption{Kinematic parameters of legged and legless jumpers.}
		\label{tab:jump}  
	\end{table*}
	
	Notwithstanding their drastically different morphologies, froghoppers and the larvae of the fruit fly and gall midge have convergently evolved to jump with a take-off angle close to 60$^\circ$. The understanding of the reasons behind this peculiar choice could shed new light on the selective pressure exerted by the geometrical and statistical features of the environment, and drive the development of bio-inspired robotic devices suitable for efficient exploration of territories with unknown features.
	Moreover, froghoppers represent the key vectors for the \textit{Xylella  fastidiosa}  Wells  bacterium, which led to a dramatic epidemic disease on olive trees in the Mediterranean area \cite{Almeida2016}. The dispersal capabilities of froghoppers are still not very well known and a recent field study has shown that they could be more effective than expected \cite{Bodino2020}. A deeper understanding of the key factors that lead to the effectiveness of dispersion of froghoppers could help the identification of strategies to mitigate the fast spread of the dieback of olive trees.
	
	In this work, we show that a take-off angle of 60$^\circ$ maximizes the probability of overcoming obstacles of random size and position scattered across the landscape. We solve numerically the dimensionless equations that describe the kinematic motion of the animal in the presence of air drag and show that the features of the motion are completely determined by the dimensionless Reynolds and Froude numbers at take-off. We demonstrate that a take-off angle of 60$^\circ$ maximizes the probability of a successful jump in a very wide region of the parameter space, largely encompassing the conditions of interest for the jump of froghoppers and the larvae of fruit fly and gall midge.

\section{Optimal strategy for jumping over a random obstacle}	
	Living beings like froghoppers and the larvae of fruit fly and gall midge have a typical size of the order of a fraction of 1 cm  (Table \ref{tab:jump}) and are surrounded by obstacles whose size can largely exceed theirs. Under these conditions, both the maximum height of the jump and its range become important in overcoming an obstacle. A leap angle of 60$^\circ$ represents a good compromise between jump height and range because it maximizes their product or, which is the same, the area below the trajectory. To show this, let us assume that the take-off of the animal occurs at a fixed velocity $v_0$, and is only affected by the gravitational acceleration $g$, in the absence of air friction. Under these conditions, the motion completely occurs in a plane perpendicular to the ground and the equations of motion $\ddot x=0$  and $\ddot y=-g$ can be easily integrated to yield the time evolution of the components of the displacement from the initial position:
	\begin{align} \label{eq:hor}
	x(t)&=v_0\,\cos(\theta)\,t,\\
	\label{eq:vert}
	y(t)&=v_0\,\sin(\theta)\, t- \frac{1}{2}\,g\, t^2,
	\end{align}
	where $x$ and $y$ are the horizontal and vertical displacements, and $\theta$ is the take-off angle.
	The range of the jump, \textit{i.e.} the distance traveled along the horizontal direction before landing at $y=0$, is:
	
	\begin{equation}
	x_R(\theta)=2\,x_M\,\sin(\theta)\,\cos(\theta),
	\label{eq:range}
	\end{equation}
	where $x_M$ is the maximum range achieved for a take-off angle of 45$^\circ$:
	\begin{equation}
	x_M={v_0}^2/g.
	\label{eq:max_range}
	\end{equation}
	
	\begin{figure}[b!]
		\centering
		\resizebox{0.45 \textwidth}{!}{
			\includegraphics {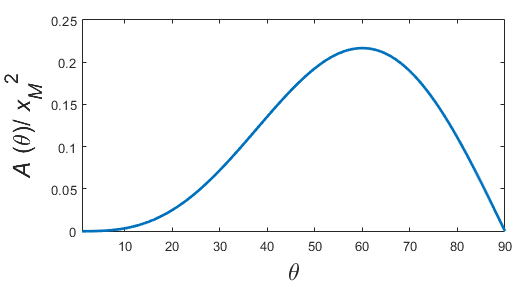}
		}
		\caption{ \textbf{Area below the trajectory as a function of take-off angle.} The area exhibits an absolute maximum at $\theta=60^\circ$. 
		}
		\label{fig:area}
	\end{figure}
	By integrating the trajectory across $x$ from the initial position $x=0$ to the range of the jump $x_R$  the area below the curve as a function of $\theta$ (Fig. \ref{fig:area}) is:
	\begin{equation}
	A(\theta)=\int_{0}^{x_R} y(x) \, d x= \frac{2}{3}\,{x_M}^2 \sin^3(\theta)\,\cos(\theta)
	\label{eq:area}
	\end{equation}
	and derivation of Eqn. \ref{eq:area} with respect to $\theta$ immediately shows that the area is has an absolute maximum for $\theta=60^\circ$.
	From Eqn. \ref{eq:area} one can appreciate that the area below the trajectory is proportional to the product between the range (Eqn. \ref{eq:range}) and top height $h(\theta)=x_M\sin^2(\theta)/2$ of the jump.
	
\begin{figure}[t]
		\centering
		\resizebox{0.45 \textwidth}{!}{
			\includegraphics {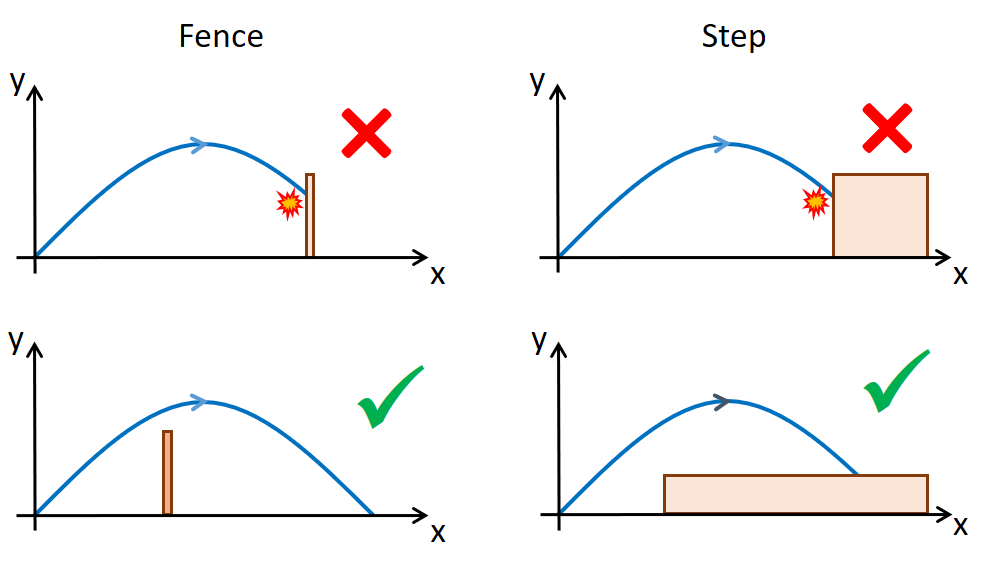}
		}
		
		\caption{ \textbf{Success of a jump.} Obstacles can be either represented by vertical fences (left column) or steps (right column). If the top-left edge of the obstacle is located above the trajectory the jump fails (top row), while when it is below the jump is successful (bottom row). 
		}
		\label{fig:obstacles}
	\end{figure}

	Although the maximization of the area below the trajectory optimizes both the range and the maximum height of the jump, one might argue that this combination of the two parameters is arbitrary and does not lead necessarily to optimal performances in a natural landscape. 
	Conversely, we will demonstrate that in a landscape made of obstacles of random size and position the probability of overcoming an obstacle with a jump is directly proportional to the area below the trajectory.
	Let us assume that \textit{i)} the landscape is populated by obstacles, which can be either vertical fences (Fig. \ref{fig:obstacles}, left column), or steps (Fig. \ref{fig:obstacles}, right column); \textit{ii)} a jump is considered not successful when the trajectory passes below the top left edge of an obstacle (Fig. \ref{fig:obstacles}, top row), and successful when it passes above it (Fig. \ref{fig:obstacles}, bottom row) \textit{iii)} the height of the obstacles is uniformly distributed in the range $\left[0,H \right] $, where $H>x_M/2$, and their position is uniformly distributed in the range $\left[0,L \right] $, where $L>x_M$; \textit{iv)} the initial velocity $v_0$ of the jump is fixed. 
	
		\begin{figure}[b]
		\centering
		\resizebox{0.45 \textwidth}{!}{
			\includegraphics {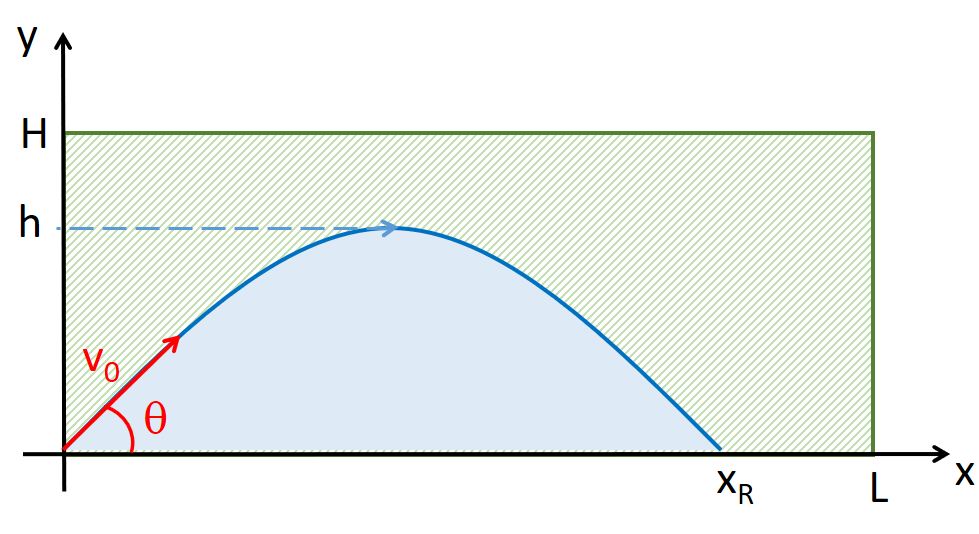}
		}
		
		\caption{ \textbf{Probability of a successful jump.} The solid light-blue region below the trajectory marks the regions of possible positions of the top edge of the obstacle in a successful jump, while the green rectangle all the possible positions. 
		}
		\label{fig:prob}
	\end{figure}
	
	Due to the uniform distribution of height and position of the obstacles, the probability $P(\theta)$ of a successful jump can be calculated directly as the area of the region of possible positions of the top-left edge of the obstacle  below the trajectory (Fig. \ref{fig:prob}, solid light-blue area below the trajectory), normalized by the area of the region of all the possible positions (Fig. \ref{fig:prob}, rectangle of base $L$ and height $H$):
	\begin{equation}
	P(\theta)=\frac{A(\theta)}{H\,L}.
	\label{eq:prob}
	\end{equation}
	Equation \ref{eq:prob} shows that the area $A(\theta)$ below the trajectory represents a direct measure of the probability of a successful jump and, in combination with Eqn. \ref{eq:area}, that the maximum probability is achieved for a take-off angle of $60^\circ$ (Fig. \ref{fig:area}).
	
	\section{Effect of air friction}
	The results derived in the previous section are obtained under the implicit hypothesis of negligible air friction. However, as discussed by Vogel \cite{Vogel2005}, the bio-ballistics of small projectiles like the insects listed in Table 1 is strongly affected by air drag, which becomes increasingly important as the size of the projectile is diminished. Under such circumstances, the presence of air drag can determine a significant decrease of both the range and height of the jump and in turn of the area below the trajectory. Indeed, Eqn. \ref{eq:area} for the area below the trajectory has been obtained in the absence of air friction.
	We will show that, although air drag affects significantly the trajectories of the insects listed in Table 1, the optimal take-off angle is affected only marginally by its presence.
	One firm result of the model reported above is represented by Eqn. \ref{eq:prob}, which states that the probability of a successful jump is proportional to the area under the trajectory.  This result is valid under generic conditions, both in the presence and in the absence of air drag. Therefore, to calculate the probability of a successful jump under realistic conditions one just needs to determine the area below the trajectories in the presence of air drag. The effect of inertia and viscous drag on the body are combined into the dimensionless Reynolds number:
	\begin{equation}
	{\mathrm {Re}}= \frac{\rho\, l \, v}{\eta},
	\label{eq:Re}
	\end{equation}
	
	where $\rho$ and $\eta$ are respectively the density of air (1.2 kg/m$^3$ at 20$^\circ$C) and its shear viscosity (1.8$\times10^{-5}$  Pa s), and $l$ is a typical effective diameter of the body (Table \ref{tab:jump}). When ${\mathrm {Re}}<1$ the motion of the body occurs in the Stokes regime, and the drag force is proportional to $v$, while for Re$>$1000 the flow occurs in the Newton regime, and the drag force becomes proportional to $v^2$. Following Vogel \cite{Vogel2005}, we introduce a drag coefficient that takes into account the transition between these regimes:
	\begin{equation}
	C_d\left({\mathrm {Re}}\right)=\frac{24}{{\mathrm {Re}}}+\frac{6}{1+{\mathrm {Re}}^{1/2}}+0.4.
	\label{eq:Cd}
	\end{equation}
	The modulus of the drag force acting of the body can be calculated as
	\begin{equation}
	D=\frac{1}{2}C_d\, \rho \, S \,v^2,
	\label{eq:drag}
	\end{equation}
	where $S=\pi(l/2)^2$ is the cross sectional area that the body offers to the air flow.
	Following Landau and Lifschitz \cite{Landau1987Fluid}, gravitational effects can be accounted for by introducing the dimensionless Froude number:
	\begin{equation}
	{\mathrm {Fr}}= \frac{v^2}{l\,g'\,},
	\label{eq:Fr}
	\end{equation}
	where $g'=g\left( \rho_p-\rho\right)/\rho$ is the reduced acceleration of gravity \cite{lee2012turbulent} and $\rho_p$ is density of the insect. 
	
\begin{figure}[t]
		\centering
		\resizebox{0.48 \textwidth}{!}{
			\includegraphics {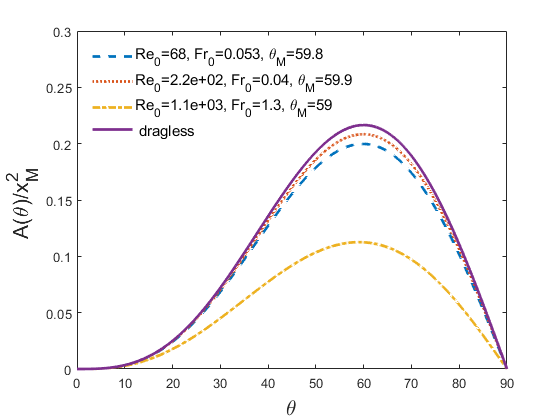}
		}
		
		\caption{ \textbf{Numerically calculated area below the trajectories in the presence of air drag.}  dashed line: gall midge; dotted line: fruit fly larva; dashed-dotted line: froghopper. The parameters used to process the trajectories mirror those reported in Table 1. The solid line represents the area in the absence of air drag, calculated from Eqn. \ref{eq:area}.  
		}
		\label{fig:friction}
	\end{figure}

	The equations of motion can be written in terms of the dimensionless variables  $(\tilde{x},\tilde{y})=(x,y)/x_M$ and $\tilde{t}=t/t_0$ , where $t_0=v_0/g$ is a characteristic time of flight:  
	\begin{align} 
	\label{eq:drag_hor}
	\ddot{\tilde{x}}&=-\mu \, \dot{\tilde{x}}\\
	\label{eq:drag_vert}
	\ddot{ \tilde{y}}&=-\mu \, \dot{\tilde{y}}-1,
	\end{align}
	where
	\begin{equation} 
	\label{eq:mu}
	\mu=\frac{3}{4}C_d\left({{\mathrm {Re}}_0}\cdot\tilde{v}\right)\cdot{{\mathrm{Fr}}_0}\cdot\tilde{v}
	\end{equation}
	is a dimensionless drag coefficient that only depends on the dimensionless velocity $\tilde{v}$ and on the Reynolds and Froude numbers at take-off, defined by  ${\mathrm {Re}}_0=\rho l v_0 / \eta$ and ${\mathrm{Fr}}_0= {v_0}^2 /l g'$, respectively. According to Eqns. \ref{eq:Cd} and \ref{eq:mu},
	for small ${\mathrm {Re}}_0$, the drag coefficient attains a constant value $\mu_0\sim 18\,{\mathrm{Fr}}_0 /\mathrm{Re}_0$ and Eqns. \ref{eq:drag_hor} and \ref{eq:drag_vert} can be solved analytically. For larger ${\mathrm {Re}}_0$, the nonlinear dependence of $\mu$ from $\tilde{v}$ makes the equation hard to solve analytically, but solutions can be still found numerically. We integrated the equations numerically by using the Euler's method detailed in \cite{vogel1988} to obtain the area under the trajectory as a function of take-off angle for a set of parameters mirroring the three representative cases of leaping insects and larvae detailed in Table 1. The numerically calculated dimensionless areas $\tilde{A}(\theta)$ are shown in Fig. \ref{fig:friction}, where dashed and dotted lines represent results obtained in the presence of air drag, while the solid line those obtained without air drag. One can appreciate that, although air drag determines a decrease of $\tilde{A}(\theta)$, the maxima of the curves are located in a narrow range of angles $59^\circ<\theta<59.9^\circ$, very close to the value $\theta=60^\circ$ predicted analytically in the absence of air drag. 
	
\begin{figure}[b]
		\centering
		\resizebox{0.48 \textwidth}{!}{
			\includegraphics {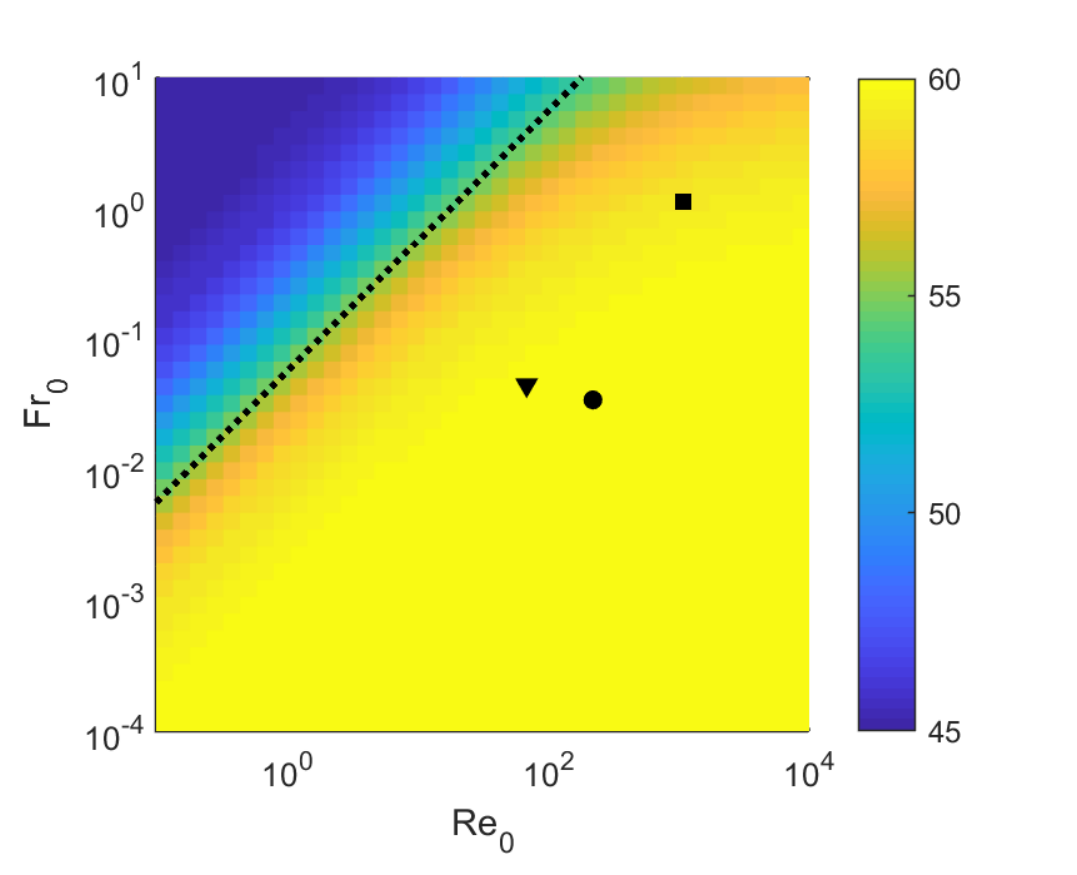}
		}
		\caption{ \textbf{Optimal leap angle as a function of the Reynolds and Froude numbers  at take-off.}  Black symbols:
			gall midge (downward triangle);  fruit fly larva (circle); froghopper (square). The dotted line corresponds to ${\mathrm{Fr}}_0={\mathrm {Re}}_0 / 18$, which is equivalent to the condition $\mu=1$ in the low Reynolds number limit (see Eqns. \ref{eq:Cd} and \ref{eq:mu})}
		\label{fig:friction_Re_Fr}
	\end{figure}

	To assess the robustness of these results, we systematically investigated the optimal leap angle,  corresponding to the maximum of $\tilde{A}(\theta)$, over a wide range of the two control parameters ${\mathrm {Re}}_0$ and ${\mathrm{Fr}}_0$. As shown in Fig. \ref{fig:friction_Re_Fr}, the parameters of all the jumpers considered in this work fall well within a large domain of the parameter space where the optimal leap angle is almost indistinguishable from the drag-free ideal case $\theta=60^\circ$. For small Reynolds numbers (${\mathrm {Re}}\lesssim 10^2$) this domain is identified by the simple condition $\mu\ll1$, which corresponds to ${\mathrm{Fr}}_0\ll {\mathrm {Re}}_0/18$ (dotted line in Fig. \ref{fig:friction_Re_Fr}).
	
\section{Discussion}
	We have demonstrated that the fact that a take-off angle close to $\theta=60^\circ$ maximizes the probability of success of a jump in the presence of uniformly distributed random obstacles represents a robust result that does not depend on the details of the model adopted. This remarkable result can be profitably used to optimize the features of autonomous robots used for the exploration of environments populated by unknown obstacles, such as other planets \cite{Burdick}, or Earth regions than can be dangerous for human beings, like nuclear disaster sites or earthquakes’ scenarios \cite{Pratt, murphy}. A well-known approach is to use rover robots that have usually large size and mass, move by using wheels or tracks, and navigate thanks to a high degree of technology and a closed-loop control of their movements. Although these characteristics allow a fine control of the navigation, they accomplish a limited capability of mapping extended territories.  A different approach is the use of colonies of small and agile robots, typically tens \cite{Zhakypov}, with simple design and functions and open-loop control, which can map extended territories.
	As the size of robots decreases, they likely have to overcome obstacles whose size is comparable or even larger than their own one \cite{Armour07,Armour10}. A bio-mimetic approach suggests that jumping has a great potentiality of success \cite{Zhang17,Ribak,Zhang20} in rough terrains. In fact, in recent years a large number of researches has focused on the refinement of miniaturized robots inspired to jumping organisms, like for example froghoppers \cite{jung16}, locusts \cite{Zaitsev}, insects \cite{Truong,Ribak,Noh}, or even soft worms \cite{Ahn,Tolley}. 
	The final choices made in the design of a jumping robot result from a complex balance among constraints connected to take-off, air flight, and landing. Flight issues have often to do with posture adjustment, landing with stability and take-off with the mechanisms of energy storing and fast conversion in kinetic energy for jumping \cite{jung16,Ahn,Zaitsev}. To our knowledge, the probability of success in overcoming an obstacle of random size and position has not been considered yet during the design process, and the take-off angle is often chosen \textit{a priori}  \cite{Kovac,Ahn}.
	
	The results reported in this work could inspire a different approach in the design of miniaturized robots, which could be particularly effective for the challenging case of groups of small robots that collectively explore a rough terrain. Under these conditions the implementation of a  robot hardwired to jump at an angle of $60^\circ$ would allow attaining optimal performances in the exploration of unknown rough regions, sided by an extremely simple conceptual design.
\section*{Data accessibility}	
The Matlab code used for the numerical integration of the equations of motion in the presence of air drag is provided as Electronic Supporting Material.
\\
\section*{Authors' contributions}	
FG designed the study, developed the model, performed numerical analysis and helped draft the manuscript; SS conceived the study, contributed to the development of the model and critically revised the manuscript; MC participated in the design of the study and helped draft the manuscript; AV conceived the study, designed the study, coordinated the study, contributed to the development of the model, performed numerical analysis and drafted the manuscript. All authors gave final approval for publication and agree to be held accountable for the work performed therein.

\section*{Acknowledgements}
We thank Marcello Re for valuable comments and suggestions.

\bibliography{jump}{}

\begin{thebibliography}{10}

\bibitem{biewener2018}
Biewener AA, Patek SN.
\newblock Animal locomotion.
\newblock Oxford, United Kingdom: Oxford University Press; 2018.

\bibitem{irschick2016}
Irschick DJ, Higham TE.
\newblock Animal athletes: an ecological and evolutionary approach.
\newblock Oxford New York: Oxford University Press; 2016.

\bibitem{Toro2004}
Toro E, Herrel A, Irschick D.
\newblock The Evolution of Jumping Performance in {CaribbeanAnolisLizards}:
  Solutions to Biomechanical Trade-Offs.
\newblock Am Nat. 2004 Jun;163(6):844--856.
\newblock doi:10.1086/386347.

\bibitem{Patek2011}
Patek SN, Dudek DM, Rosario MV.
\newblock From bouncy legs to poisoned arrows: elastic movements in
  invertebrates.
\newblock J Exp Biol. 2011 May;214(12):1973--1980.
\newblock doi:10.1242/jeb.038596.

\bibitem{Burrows2008}
Burrows M, Shaw SR, Sutton GP.
\newblock Resilin and chitinous cuticle form a composite structure for energy
  storage in jumping by froghopper insects.
\newblock BMC Biol,. 2008 Sep;6(1).
\newblock doi:10.1186/1741-7007-6-41.

\bibitem{Burrows2003}
Burrows M.
\newblock Froghopper insects leap to new heights.
\newblock Nature. 2003 Jul;424(6948):509--509.
\newblock doi:10.1038/424509a.

\bibitem{Maitland1992}
Maitland DP.
\newblock Locomotion by jumping in the Mediterranean fruit-fly larva Ceratitis
  capitata.
\newblock Nature. 1992 Jan;355(6356):159--161.
\newblock doi:10.1038/355159a0.

\bibitem{Kier2012}
Kier WM.
\newblock The diversity of hydrostatic skeletons.
\newblock J Exp Biol. 2012 Mar;215(8):1247--1257.
\newblock doi:10.1242/jeb.056549.

\bibitem{Farley2019}
Farley GM, Wise MJ, Harrison JS, Sutton GP, Kuo C, Patek SN.
\newblock Adhesive latching and legless leaping in small, worm-like insect
  larvae.
\newblock J Exp Biol. 2019 Aug;222(15):jeb201129.
\newblock doi:10.1242/jeb.201129.

\bibitem{Goetzke2019}
Goetzke HH, Pattrick JG, Federle W.
\newblock Froghoppers jump from smooth plant surfaces by piercing them with
  sharp spines.
\newblock Proc Natl Acad Sci USA. 2019 Feb;116(8):3012--3017.
\newblock doi:10.1073/pnas.1814183116.

\bibitem{Almeida2016}
Almeida RPP.
\newblock Can Apulia{\textquotesingle}s olive trees be saved?
\newblock Science. 2016 Jul;353(6297):346--348.
\newblock doi:10.1126/science.aaf9710.

\bibitem{Bodino2020}
Bodino N, Cavalieri V, Dongiovanni C, Simonetto A, Saladini MA, Plazio E,
  et~al.
\newblock Dispersal of Philaenus spumarius (Hemiptera: Aphrophoridae), a Vector
  of Xylella fastidiosa, in Olive Grove and Meadow Agroecosystems.
\newblock Environmental Entomology. 2020 Dec.
\newblock doi:10.1093/ee/nvaa140.

\bibitem{Vogel2005}
Vogel S.
\newblock Living in a physical world {II}. The bio-ballistics of small
  projectiles.
\newblock J Biosci. 2005 Mar;30(2):167--175.
\newblock doi:10.1007/bf02703696.

\bibitem{Landau1987Fluid}
Landau LD, Lifshitz EM.
\newblock {Fluid Mechanics}. vol.~6 of Course of Theoretical Physics.
\newblock 2nd ed. Pergamon; 1987.

\bibitem{lee2012turbulent}
Lee JHw, Chu V.
\newblock Turbulent jets and plumes: a Lagrangian approach.
\newblock Springer Science \& Business Media; 2012.

\bibitem{vogel1988}
Vogel S.
\newblock Life's devices: the physical world of animals and plants.
\newblock Princeton, N.J: Princeton University Press; 1988.

\bibitem{Burdick}
Burdick J, Fiorini P.
\newblock Minimalist Jumping Robots for Celestial Exploration.
\newblock Int J Robot Res. 2003 July–August;22(7-8):653--674.
\newblock doi:10.1177/02783649030227013.

\bibitem{Pratt}
Pratt GA.
\newblock Robot to the rescue.
\newblock Bull At Sci. 2014;70:63--69.
\newblock doi:10.1177/0096340213516742.

\bibitem{murphy}
Murphy RR.
\newblock Disaster robotics.
\newblock Cambridge, MA, USA: MIT Press; 2014.

\bibitem{Zhakypov}
Zhakypov Z, Mori K, Hosoda K, Paik J.
\newblock Designing minimal and scalable insect-inspired multi-locomotion
  millirobots.
\newblock Nature. 2019;571:381–386.
\newblock doi:10.1038/s41586-019-1388-8.

\bibitem{Armour07}
Armour R, Paskins K, Bowyer A, Vincent J, Megill W.
\newblock Jumping robots: a biomimetic solution to locomotion across rough
  terrain.
\newblock Bioinsp Biomim. 2007;2:S65–S82.
\newblock doi:10.1088/1748-3182/2/3/S01.

\bibitem{Armour10}
Armour RH.
\newblock A Biologically Inspired Jumping and Rolling Robot.
\newblock University of Bath: PhD Thesis; 2010.

\bibitem{Zhang17}
Zhang Z, Zhao J, Chen H, D C.
\newblock A Survey of Bioinspired Jumping Robot: Takeoff, Air Posture
  Adjustment, and Landing Buffer.
\newblock Appl Bionics Biomec. 2017;2017:780160.
\newblock doi:10.1155/2017/4780160.

\bibitem{Ribak}
Ribak G.
\newblock Insect-inspired jumping robots: challenges and solutions to jump
  stability.
\newblock Curr Opin Insect Sci. 2020;42:32–38.

\bibitem{Zhang20}
Zhang C, Zou W, Ma L, Z W.
\newblock Biologically inspired jumping robots: A comprehensive review.
\newblock Rob Auton Syst. 2020;124:103362.
\newblock doi:10.1016/j.robot.2019.103362.

\bibitem{jung16}
Jung GP, Cho KJ.
\newblock Froghopper-inspired direction-changing concept for miniature jumping
  robots.
\newblock Bioinspir Biomim. 2016;11:056015.
\newblock doi:10.1088/1748-3190/11/5/056015.

\bibitem{Zaitsev}
Zaitsev V, Gvirsman O, Hanan UB, Weiss A, Ayali A, Kosa G.
\newblock A locust-inspired miniature jumping robot.
\newblock Bioinspir Biomim. 2015;10:066012.
\newblock doi:10.1088/1748-3190/10/6/066012.

\bibitem{Truong}
Truong NT, Phan HV, Park HC.
\newblock Design and demonstration of a bio-inspired flapping-wing-assisted
  jumping robot.
\newblock Bioinspir Biomim. 2019;14(3):036010.
\newblock doi:10.1088/1748-3190.

\bibitem{Noh}
Noh JS, Kim SW, An S, Koh JS, Cho KJ.
\newblock Flea-Inspired Catapult Mechanism for Miniature Jumping Robots.
\newblock IEEE Trans Robot. 2012 Oct;28(5):1007.
\newblock doi:10.1109/TRO.2012.2198510.

\bibitem{Ahn}
Ahn C, Liang X, Cai S.
\newblock Bioinspired Design of Light-Powered Crawling, Squeezing, and Jumping
  Untethered Soft Robot.
\newblock Adv Mater Technol. 2019;4:1900185.
\newblock doi:10.1002/admt.201900185.

\bibitem{Tolley}
Tolley MT, Shepherd RF, Karpelson M, Bartlett NW, Galloway KC, Wehner M, et~al.
\newblock An Untethered Jumping Soft Robot.
\newblock IEEE/RSJ International Conference on Intelligent Robots and Systems
  Chicago, IL, USA. 2014 Sep;14-18:561.

\bibitem{Kovac}
Kova\v{c} M, Fuchs M, Guignard A, Zufferey JC, Floreano D.
\newblock A miniature 7g jumping robot.
\newblock IEEE International Conference on Robotics and Automation Pasadena,
  CA, USA. 2008 May;19-23:373.

\end{thebibliography}

\end{document}